\documentstyle[aps,prb]{revtex}
\begin{document}
\draft
\title{Persistent Current in Normal Metals}
\author{Pritiraj Mohanty \\
        {\it Condensed Matter Physics 114-36, California Institute of Technology, 
        Pasadena, CA 91125\\
	(Ann. Phys. (Leipzig) {\bf 8}, 549 (1999))}  \\
}
\maketitle

\begin{abstract}
We discuss the recent experiments on persistent current in metallic rings
in the backdrop of low temperature decoherence. The observed size of
the persistent current, typically on the order of the Thouless energy, $ e/\tau_D$, 
is much larger than the theoretical
results, obtained with or without electron interaction.
In considering the phenomenology of both decoherence and persistent current,
usually observed in similar systems, we argue towards a dynamic role played by
decoherence in the generation of a persistent current. 
A field-induced phase shift from near-equilibrium high-frequency 
fluctuations---which otherwise gives rise to decoherence---under
certain conditions of periodicity and asymmetry due to disorder,
is argued to induce a steady state diffusion current on 
the order of $e/\tau_D$, 
comparable to the observed persistent current.
\end{abstract}

\section{Introduction}
One of the most spectacular aspects of the quantum mechanics of
an electron in a small conducting system is interference; This results in 
the modulation of conductance periodic in enclosed flux with a period of 
the fundamental flux quantum $h/e$--the effect 
dubbed as the Aharonov-Bohm effect in mesoscopic physics. 
Similarly, a small metallic ring threaded with an Aharonov-Bohm flux displays 
a thermodynamic persistent current, signifying quantum coherence of electrons 
in the ground state. But an unavoidable feature associated with any interference 
effect in mesoscopic system is decoherence. Inside a disordered conductor, 
the decoherence of an electron occurs due to its interaction with its environment:
coupling to localized spins, electron-phonon interactions and 
electron-electron interactions, the latter dominating at low temperature.
Though the coupling to an environment results in decoherence, it is quite feasible
for the coupling to induce a nontrivial coherent effect, 
contingent upon other requirements.

The presence of a flux $\Phi$ modifies the boundary condition for the electron
wave function in an isolated ring, requiring dependence of the 
{\it equilibrium} free energy $F$ on $\Phi$; This results in an 
{\it equilibrium} current in the ring\cite{original}: 
\begin{equation} 
I(\Phi) = -\delta F(\Phi)/\delta \Phi. 
\end{equation}
\noindent
$I(\Phi)$ is periodic in flux with fundamental periodicity 
$h/e$, the flux quantum, and only exists in the presence of a magnetic field. 
For a disordered ring, the fundamental harmonic (with 
flux periodicity $h/e$) is strongly suppressed, whereas the first harmonic (with 
flux periodicity $h/2e$) survives due to the contribution of time-reversed paths 
of the electron\cite{imry:book}. Another condition is that only those electrons whose wave 
functions are sufficiently extended to wrap around the ring carry the persistent 
current; the phase decoherence 
length is larger than the length of the ring, 
\begin{equation} 
L_\phi > L, \mbox{or equivalently,}\quad \tau_\phi > \tau_D, 
\end{equation}
\noindent
where  $\tau_\phi = L_\phi^2/D$, and the time for the diffusion of the electron
around the ring, $\tau_D = L^2/D$; The current $I(\Phi)$ is not expected to decay 
once this condition is satisfied.

Persistent current in normal metals is already known to 
exist\cite{levy1,chandrasekhar,mailly,persistent}, though verified 
only in a handful of experiments. Recent experimental results on the size
both in the $h/e$ and $h/2e$ components are about two orders of magnitude larger 
than the most hopeful theoretical result, obtained with the inclusion of electron 
interaction perturbatively\cite{ae}. The simultaneous requirement of explaining the 
observed diamagnetic sign of the $h/2e$ component makes the comparison with 
theory even worse. Without electron interaction, the size of the current is too 
small, with the wrong sign\cite{non-interacting}. 

Keeping the inadequecy of numerous attempts in view\cite{ae,non-interacting,gefen,schwab}, 
and motivated by additional experimental observations, we explore a possible 
connection between the observed 
persistent current and decoherence\cite{prl97,prb97}. In an excursion from the 
fundamental premise of its existence as an {\it equilibrium} thermodynamic 
property, we argue that the observed current in the experiments may {\it not} 
be the textbook persistent current--an aspect of complete quantum coherence 
of electrons in the ground state {\it in equilibrium}. In stead it could 
perhaps be a steady state dc current 
generated out of coherent phase shifts due to the presence of near-equilibrium or 
non-equilibrium fluctuations. Our argument is motivated by the persistence of 
decoherence at low temperature\cite{prl97} in the range where the persistent current is 
measured\cite{levy1,chandrasekhar,persistent,levy2}. 
Assumption of the low temperature decoherence implies the presence of 
time-dependent field fluctuations. These fluctuations induce random phase shifts 
in the electron wave function, resulting in decoherence upon averaging. In an 
isolated {\it periodic} structure, correlation can arise from these random 
field-induced phase shifts for frequencies of the field matching with the 
traversal frequency of the electron diffusion around the length of the periodic 
structure, i.e. for $\omega = 2\pi/\tau_D$. For these paths, the phase shift
in the electron wave function induced by the field fluctuations is $2\pi$;
The electrons can, in principle, carry energy from the field fluctuations, though
a steady state finite current can only be maintained by the electrons in 
these paths due to the absence of decoherence. Other paths, or equivalently,
frequencies higher or lower than $1/\tau_D$ contribute to a transient current 
which dies down within a time scale of $1/\tau_\phi$.

In the next section we report our experimental results on the first complete
measurement of the persistent current, including comparisons
with previous experiments. In the subsequent section we briefly review the 
previous theoretical attempts with and without electron interaction. 
Finally, we discuss the plausible dynamic role played by an environment of 
high-frequency electromagnetic field fluctuations. The environment is assumed 
to be arising from electron interaction perhaps in the strong 
coupling limit, though such an assumption is not necessary.

\section{Phenomenology}

Persistent current in normal metal systems 
has been observed only in a handful of
experiments\cite{levy1,chandrasekhar,persistent}. The magnitude of the $h/e$ 
current along with its temperature 
dependence has been measured in experiments on single Au 
rings\cite{chandrasekhar}. The $h/e$ component of the current has also been
observed in a semiconducting GaAs-AlGaAs heterostructure ring\cite{mailly}. 
The $h/2e$ current along with its temperature dependence has been measured, 
in the first persistent current experiment\cite{levy1}, in an array of 10 million 
Cu rings. The $h/e$ component measured in the ballistic ring\cite{mailly} has 
the expected magnitude of $ev_F/L$, where $v_F$ is the Fermi velocity. The size of
the $h/e$ current measured in single Au rings\cite{chandrasekhar} is two orders 
of magnitude larger than the anticipated value of $ev_F/L (l_e/L)$ for the typical 
current in a single disordered ring; $l_e \sim D/ v_F$ is the elastic mean free path
denoting disorder. Likewise, the $h/2e$ current measured in the multiple Cu ring 
experiment was more than two orders of magnitude larger in size than the 
anticipated values of $\sim e\Delta/\hbar$ for the  $h/2e$ current averaged over
disorder and many rings. The sign of the $h/e$ current was determined to be
random as expected, whereas the sign of the $h/2e$ was not determined. Temperature
dependence of both the components of current was found to be exponentially decaying
with increasing temperature.  
 
In what follows, we describe the results of our recent experiment without going 
through the experimental details, published elsewhere\cite{persistent}.
We have measured the magnitude, temperature dependence,
and sign near zero field of both the $h/e$-- and $h/2e$--periodic components 
of the persistent current in an array of 30 Au rings. Let us first summarize the 
results.

\subsection{The $h/2e$ current} 
(a) The magnitude of $\sim 0.5 e/\tau_D$  observed in the 
$h/2e$ channel is in agreement with the earlier experiment on Cu rings.
(b) Temperature dependence of $e^{-T/89\mbox{mK}}$, 
though within a (limited) range of 10-150 mK, is comparable to 
previous measurements. (c) The sign of the
current near zero field was determined to be diamagnetic. (d) We have observed a strong
magnetic field dependence on all our $h/2e$ current traces  with a characteristic 
field $H_c = \sqrt{3}\hbar/ewL$, where $w$ is the width of the ring. 
This field dependence is perhaps due to the magnetic field penetration into the arms of 
the ring corresponding to a decay rate $1/\tau_H = D(eHw/\sqrt{3}\hbar)^2$, 
comparable to magnetic-field-induced phase-breaking in weak localization.

\subsection{The $h/e$ current} 
(a)  We observe a magnitude of 
$2.3 eE_c/\hbar \simeq 10^{-2} (ev_F/L)$ for the average $h/e$ current. 
Unfortunately, a comparison with earlier experiments is not possible, since
the average $h/e$  current--in an ensemble of many rings--has not been measured 
previously. (b) The temperature dependence is found to be $e^{-T/166\mbox{mK}}$. 
(c) The sign of the current near zero field is diamagnetic. (d) The $h/e$ current 
also showed a suppression of the size with increasing $\Phi$, 
though the effect was weaker with a larger $H_c$ compared to the $H_c$ of the 
$h/2e$ current.  

\begin{figure}
 \vbox to 14.0cm {\vss\hbox to 7cm
 {\hss\
   {\includegraphics{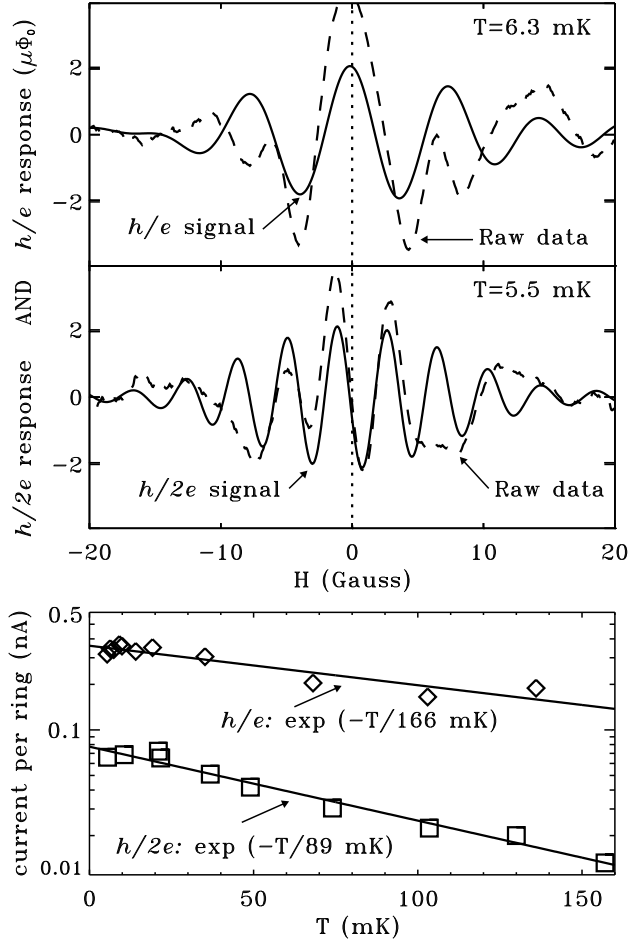}
   }
  \hss}
 }
\caption{The observed $h/e-$ and $h/2e-$flux periodic currents are displayed 
in the top panel. The bottom panel shows the temperature dependence of both
components.}
\end{figure}

\subsection{Sample parameters}

The Au rings are described by the following parameters: $2.56 \pm 0.05 \mu m$
in diameter, $120 \pm 20 nm$ wide, and $60 \pm 2 nm$ thick. Their resistance per
square of $ 0.15 \Omega/\mbox{square}$ corresponds to a diffusion constant of 
$D = v_F l_e/2 = 0.06 m^2/s$, with an elastic mean free path of $l_e \simeq 87 nm$.
These parameters as well as $L_\phi$ of $16 \mu m$ are 
obtained in a separate transport 
measurement of weak localization. The control sample
used in this measurement 
is a $207 \mu m$ long meander with the
same thickness
and linewidth as the rings, fabricated simultaneously. 
$L_\phi$, or equivalently $\tau_\phi$, was found to be essentially temperature 
independent below 500 mK. Various energy scales of the disordered rings are the 
following: Thouless energy $E_c \sim \hbar/\tau_D = 7.3 mK$,
mean energy level spacing $\Delta \sim 1/2{\cal N}_0{\cal V} = 19 \mu K$, where
${\cal N}_0$ is the density of states, and ${\cal V}$ is the volume of each ring.

\subsection{Comparison with previous theories}

(i)The observed magnitude of the $h/2e$ current of $\sim 0.5 eE_c/\hbar$, along with
the diamagnetic sign is hard to understand in previous 
theories\cite{ae,non-interacting}; This result is
much larger than the expected average current of $e\Delta/\hbar$ in non-interacting 
calculations\cite{non-interacting}, though a typical current on the order 
of $eE_c/\hbar$ is expected.
The first theoretical calculation that includes the electron-electron 
interaction\cite{ae}
obtains a current within an order of magnitude of $eE_c/\hbar$, but it requires 
a repulsive Coulomb interaction, necessarily resulting in a paramagnetic current. 
In order to obtain a diamagnetic

\begin{figure}
 \vbox to 11.00cm {\vss\hbox to 7cm
 {\hss\
   {\includegraphics{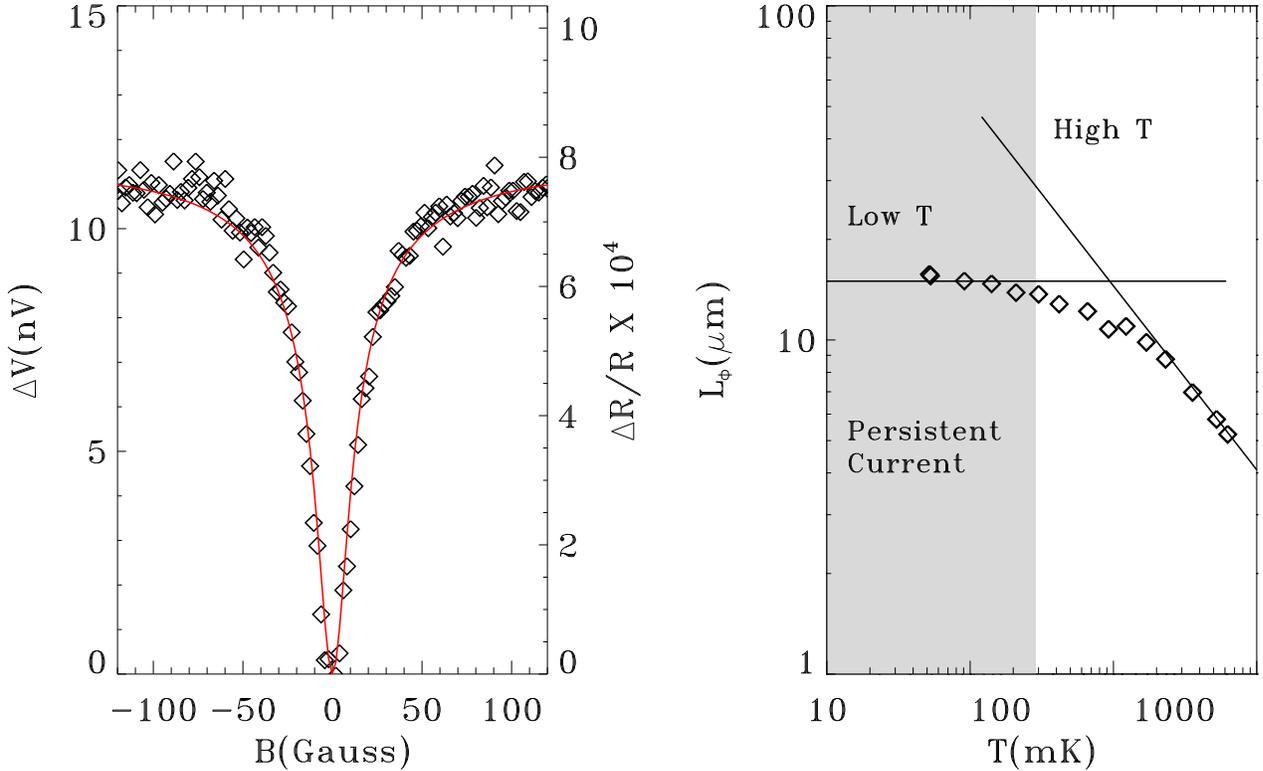}
   }
  \hss}
 }
\caption{(a)Weak localization measurement for the determination of $L_\phi$.
(b) Temperature dependence of $L_\phi$ in the corresponding control sample 
of quasi-1D Au wire. The shaded area represents the range of temperature in 
which the persistent current was measured.
}
\end{figure}
\noindent
current in such a theory, a phonon-mediated
attractive electron-electron interaction must be used; however, the small strength
of the potential of such an interaction gives a result even smaller by another order 
of magnitude.
(ii)
We observe a magnitude of $2.3\,eE_c/\hbar \simeq 10^{-2} (ev_F/L)$ for the
$h/e$ current in our ensemble of rings; in contrast, both non-interacting and 
interacting theories find that the average $h/e$ current should be vanishingly 
small in the diffusive limit. To avoid any confusion, we note here that our 
measurement of the {\it average} $h/e$-periodic response from thirty rings 
is smaller than the magnitude $ev_f/L$ of the {\it typical} $h/e$ current 
measured in single rings\cite{chandrasekhar}. For the typical current in a single 
ring, theory suggests an amplitude $eE_c/\hbar$, again much smaller than 
measured \cite{chandrasekhar}. In either single\cite{chandrasekhar} or multiple 
ring experiments\cite{persistent}, the measured amplitudes of the $h/e$ response 
are much larger than accounted for by any theory.                        
\noindent
(iii)The exponential dependence for both $h/e$ and $h/2e$ components on temperature
is similar to what was found in earlier experiments, in both single 
\cite{chandrasekhar} and multiple rings\cite{levy1}. Even though some 
theories\cite{ae,non-interacting} contain a temperature dependence, one must be 
cautious in comparing our result to such predictions over a limited range of very 
low temperatures.
\noindent
(iv)
Another interesting feature of our experiment, common with previous experiments, 
that decoherence rate $1/\tau_\phi$ is completely 
saturated\cite{levy2,chandrasekhar,persistent}, in the range of 
10-150 mK in which the persistent current is measured in the Au rings. 

\section{Magnitude and sign of the current in previous theories}

\subsection{Single electron picture}

In order to understand all the experimentally observed features of the 
persistent current consistently, it is important to briefly review 
the earlier attempts. The simplest model for persistent current
from an isolated system of electrons in a disordered potential involves
the modification by the flux $\Phi$ of the single-electron energy 
spectrum--neglecting the interaction between the 
electrons\cite{non-interacting,gefen}. Naively,
though incorrectly, it can be assumed that the thermodynamics of the 
electron system is governed only by the mean spacing $\Delta$ of energy 
levels of the electron eigenstates. The flux dependence of the total energy
$E(\Phi)$ can be assessed as the sum of all ocupied single-particle levels.
Since $\partial E_i(\Phi)/\partial \Phi$ alternates in sign for 
consecutive levels, cancellation in the sum leads to a current defined
by the last term around the Fermi energy $E_F$:
\begin{equation}
I(\Phi) = \sum_i {\partial E_i(\Phi) \over \partial \Phi} \sim {\Delta \over {h/e}}.
\end{equation}
\noindent
This is a strange result, since the level spacing is not affected by 
disorder($D$). Interference corrections such as weak localization modify $D$, 
leaving $\Delta$ unaffected. Thus in the single electron picture thermodynamics
of the ring is disorder independent
and is not due to quantum interference. It can be argued that thermodynamic
properties of small systems at low temperature are governed by the
mesoscopic fluctuations rather than the equilibrium distribution of the 
single-electron energy levels near $E_F$. But fluctuations in the distribution
give negligible contribution to any thermodynamic quantity in an ensemble where
the chemical potential $\mu$ is fixed. In a canonical ensemble with the number
of electrons $N$ fixed, however, non-negligible corrections to the mean value
of thermodynamic quanitities such as the persistent current.  
A trick is used, to allow for the calculation of the current by conventional
methods;
\begin{equation}
\langle ({\partial F(N,\Phi)\over  \partial \Phi})_N \rangle \equiv {\Delta \over 2}
 {\partial \over \partial \Phi} \langle (\delta N)^2\rangle_{\mu = \langle \mu \rangle},
\end{equation}
\noindent
which signifies an elegant identity that the current as the flux 
derivative of the free energy in a canonical ensemble is equivalent to 
the flux deivative of the variance in the number fluctuation 
$\langle (\delta N)^2 \rangle$
in the corresponding grand canonical ensemble. The number fluctuation in a disordered
conductor has been studied in great detail, and has been found in the presence of
a flux $\Phi$--which breaks the time-reversal symmetry--to be
\begin{equation}
\langle (\delta N(E))^2 \rangle = {4 \over \pi^2}[\ln {E \over \Delta} + \ln {E\Phi_0^2
\over E_c \Phi^2}],
\end{equation}
\noindent
valid for $\delta < E_c (\Phi/\Phi_0)^2 < E_c$. Using this one obtains an expression for
various harmonics of the current $\langle I(\Phi)\rangle = \sum_m I_m e^{2im\Phi/\Phi_0}$,
periodic in flux:
\begin{equation}
I_m = {4i\Delta \over \pi \Phi_0} e^{-2m/\sqrt{E_c \tau_\phi}} sgn(m). 
\end{equation}
\noindent
The current, obtained in this calculation, is weakly dependent on
disorder, through an exponential factor; the same factor introduces the temperature
dependence as well by the replcement of $\tau_\phi$ by $\hbar/k_BT$ at finite $T$.
The maximum value of the current $4\Delta/\pi \Phi_0 \sim e\Delta/\hbar$ remarkably
depends, once again, on the mean energy level spacing $\Delta$ of the single-electron 
eigen states. The current is observed to be paramagnetic near zero flux in this
model. Experimental observation of a current of $eE_c/\hbar = e/\tau_D$, more 
than two orders of magnitude larger than $e \Delta/\hbar$, and the observed sign are
contrary to the expectation. In the presence of strong spin-orbit scattering, as in
Au rings, the expression for the $\langle (\delta N(E))^2 \rangle$ is slightly
modified, and the sign of the current remains unchanged, failing on both counts 
to explain the experimental results.

\subsection{Inclusion of electron interaction}

In view of the experimental results it is imperative to consider 
interaction among the electrons\cite{ae}. 
Coulomb interaction among electrons give an orbital magnetic response\cite{imry:book} 
which is paramagnetic usually for a normal metal with repulsive interaction and 
diamagnetic for an attractive effective interaction--analogous to superconducting
fluctuations in the normal state at a temperature slightly above $T_c$. In
the first order perturbation theory, the interaction-induced persistent current\cite{ae}
is obtained to be
\begin{equation}
I_m \sim (e/\hbar) \lambda_c(T) E_c,
\end{equation} 
\noindent
which effectively replaces the mean level spacing, 
$\Delta \rightarrow \lambda_c(T) E_c$. The coupling constant $\lambda_c(T)$ is
related to the bare coupling constant: 
$\lambda_c(T)^{-1} = \lambda_0^{-1} + \ln (\epsilon_c/k_BT)$, and the cut-off energy 
$\epsilon_c$ is typically $E_F$ or the Debye temperature. Using $\lambda_c \simeq 0.08$
for Au, one obtains a current of an order of magnitude smaller than the experimental
value. The perturbative treatment of the Coulomb interaction thus fails to explain the
experimental results. In the absence of a nonperturbative analysis of the 
interaction-induced current, a different view towards the problem is required.

\section{Dynamic role of the electron-field coupling in persistent current}

Phase coherence is an important consideration in the existence of 
persistent current. Even inside an isolated loop, an electron possesses a
finite decoherence time $\tau_\phi$, often slightly longer than the time
taken for the electron to diffuse around the ring, $\tau_D$. 
A question arises as to how these short timescales 
$\tau_\phi$ and $\tau_D$ reconcile with the timescale of persistence, almost 
infinite. It is normally assumed, {\it ad hoc}, that the current, the flux 
derivative of the appropriate energy $I(\Phi) = -\partial F/\partial \Phi$, 
in a phase coherent ring with $L > L_\phi$ reflects absolute quantum 
coherence of each state. This conceptual 
assumption in some sense implies the absence of decoherence, or 
$1/\tau_\phi \rightarrow 0$.

Let us consider a specific decohering environment and analyze its role towards
the generation of a current\cite{kravtsov}. Though the following 
analysis is valid for any high-frequency environment\cite{imry,ralph}, we 
are interested specifically in the 
electromagnetic field\cite{aak,landau} created by the presence of other 
moving electrons, described by fluctuations with time-dependent correlation 
function. If the number of relevant degrees of freedom describing this field 
is small, thermodynamically speaking, or if the coupling with the electron 
is strong, then this decohering environment cannot be described as a true 
thermodynamic bath. This is because of competing 
timescales: Switching on the coupling between the electron and the environment 
disturbs the equilibrium of the latter. The fluctuations, described by time-dependent 
correlations, die down on a relaxation timescale which may not necessarily be 
short compared to $\tau_\phi$ or $\tau_D$, in which case the electron experiences 
truly non- or near-equilibrium fluctuations.

The time-dependent field induces a phase shift in the electron's wave function.
In order to study the phase shift due to an electromagnetic
potential in the case
of a pair of time-reversed paths ($h/2e$) or any two interfering paths (or $h/e$)
during a period
of interaction in a temporal interval of $[t_i, t_f]$, the time
dependence is defined in terms of sum and difference 
coordinates\cite{chakravarty}, $\overline{t}=(t_f+t_i)/2 ; t_0=t_f-t_i$, and thus
$t=\overline{t}+t^\prime$;
and $-t_0/2 \le t^\prime \le t_0/2$.
For an equivalent pair of time-reversed paths, $r_{{t^\prime}}^{cl}$ and
$r_{-t^\prime}^{cl}$, the phase difference due to a time-dependent potential
is given by
\begin{equation}
\delta\phi[r_{t^\prime}] = -{ie\over \hbar} \int_{-t_0/2}^{t_0/2} dt^{\prime}
[\Phi(r_{t^\prime}; \overline{t}-t^{\prime})-\Phi(r_{t^\prime}; \overline{t} +t^{\prime})],
\end{equation}
\noindent
where the time-dependent electromagnetic potential $\Phi = \int_T E(t)\dot{r}(t)dt$. 
One obtains for the phase shift:
\begin{equation}
\delta\phi[r_{t^\prime}] = -{ie\over \hbar} \int_{-t_0/2}^{t_0/2} dt
\int_{-t_0/2}^{t}  dt^{\prime}
E(r_{t^\prime};\overline{t}-t^{\prime}) (|{\dot{r}}_1(t^\prime)| - |{\dot{r}}_2(t^\prime)|.
\label{velocity}
\end{equation}                      
\noindent
An associated current $j$ can be defined via a current density
$j = \rho (|{\dot{r}}_1| e^{i\phi_1} - |{\dot{r}}_2| e^{i\phi_2} )$, written
as $[(e^{-i\phi_1} + e^{-i\phi_2})(|{\dot{r}}_1| e^{i\phi_1} - |{\dot{r}}_2| e^{i\phi_2} ) 
+ h.c.]$ or equivalently,
\begin{equation}
j \propto (|{\dot{r}}_1| -|{\dot{r}}_2|)[\mbox{constant} + \cos (\phi_1 - \phi_2)].
\end{equation}
\noindent
The phase shift in Eq.\ref{velocity} is also proportional to the velocity difference
between the two paths; the current depends quadratically on the velocities, and thus
survives ensemble averaging:
\begin{equation}
\langle j \rangle 
= -sin (4\pi\Phi/\phi_0) \langle \delta\phi(|{\dot{r}}_1| - |{\dot{r}}_2| \rangle,
\end{equation}
\noindent
whereas the phase shift, being a random quantity, is zero upon ensemble averaging.
This simple but elegant form has been derived and analyzed 
in detail earlier\cite{kravtsov}. Naively speaking, the current depends on the
average of the mean square velocity difference. The external field, in presence of
a flux, generates a drift current in both directions. The symmetry breaking, required
for a finite mean square average, is provided by the lack of inversion symmetry in
the ring due to disorder. This asymmetry of the disorder potential is very crucial
to the generation of the current. Such non-equilibrium currents in transport
configuration  have been shown to
exist as the so-called photo-voltaic effect\cite{falko}. The degree of asymmetry is 
a random function of the electron energy and it varies on a scale of $\hbar/\tau_D$. 
Thus contributions from different correlated energy intervals of width $\hbar/\tau_D$ 
will fluctuate in sign and cancel each other, suggesting that the
total current be on the order of $(e/\hbar)(\hbar/\tau_D) = e/\tau_D$. Note the similarity
to the  single electron picture\cite{non-interacting}, but with a different energy width interval for 
cancellation. 

Effect of the high-frequency field on the electron causes random phase shifts which
on averaging give a finite decoherence rate. But for a certain frequency 
$\omega = 2\pi/\tau_D$, the phase of the electron coincides with its initial phase 
after a diffusion time $\tau_D$ around the ring. In such a case, randomization
of the phase by averaging does not apply, and for this mode the current survives
without dissipating. This is because it corresponds to a phase shift of $2\pi$ during 
each trip around the ring.  This also suggests a maximum current at $\omega = 2\pi/\tau_D$ 
for a single electron diffusing in a time $\tau_D$; the size of the current roughly equals
$e/\tau_D$. More detailed calculation\cite{kravtsov} shows that the maximum current
obtained is $\sim 0.53(e/\tau_D)$, in truly excellent agreement with the measured value
of $0.5 (e/\tau_D)$. This dc response, arising as a rectification of high frequency fluctuations, 
is a steady state current in contrast to a thermodynamic current. Conceptually, a steady
state current implies that there is no net energy transfer between the bath and the 
electron in this state, whereas an equilibrium state obeys the condition of 
detailed balance and is described by a Boltzmann distribution.  
Furthermore, the current, for a pair of time-reversed paths, is diamagnetic, also in agreement with our experiment. Though a slow temperature
dependence is anticipated in theory, and the same is observed in experiments, the limited
range of data must be taken into account for agreement.

\section{Conclusion}

We discussed the results of the experiments on persistent current in normal metal systems.
We find that both the size and the sign of the observed current can be explained as a 
steady state dc response of a single electron to a high-frequency electric field. This 
field, assumed to be intrinsic to the metallic system, is perhaps generated by the motion 
of other electrons in the system. The time dependence of fluctuations in
the near-equilibrium state, argued to be important in the strong coupling case, gives
rise to this steady state dc current. Furthermore, this steady state current is 
microscopically argued to be in a non-dephasing resonant mode. The agreement between
this theory describing a persistent dc response and the observed current in experiments 
is truly excellent.

The experiments were
done in collaboration with Prof. Richard Webb and Manher Jariwala. This essay 
on the problem of persistent current, in its relation to decoherence, has 
benefited from many conversations for which I am most thankful to  
V. Ambegaokar, H. Bouchiat, E. Buks, N. Dupuis, 
Y. Gefen, Y. Imry, V.E. Kravtsov, A. Krokhin, R. Lifshitz, D. Mailly, 
G. Montambaux, P. Schwab, J. Schwarz and R. Smith. I exclusively thank 
Prof. B. Kramer for giving me the opportunity to participate in the 
Localization99 conference.


\begin{references}

\bibitem{original}
        N. Byers, and C.N. Yang, Phys Rev. Lett. {\bf 7} (1961) 46;
        F. Bloch, Phys Rev. B {\bf 2} (1970) 109;
        M. Buttiker, Y. Imry, and R. Landauer,
        Phys. Lett. {\bf 96}A (1983) 365
\bibitem{imry:book}
	Y. Imry, {\it Introduction to Mesoscopic Physics}, Oxford University Press 1997

\bibitem{levy1} L.P. Levy {\it et al.}, Phys. Rev. Lett. {\bf 64} (1990) 2074

\bibitem{chandrasekhar}
V. Chandrasekhar {\it et al.}, Phys. Rev. Lett. {\bf 67} (1991) 3578 (1991)

\bibitem{mailly} D. Mailly {\it et al.}, Phys. Rev. Lett. {\bf 70} (1993) 2020

\bibitem{persistent} P. Mohanty {\it et al.}, to be published

\bibitem{ae} V. Ambegaokar and U. Eckern, Phys. Rev. Lett. {\bf 65} (1990) 
		381; {\bf 67} (1991) 3192(C); Europhys. Lett. {\bf 13} (1990) 733
\bibitem{non-interacting} H.F. Cheung {\it et al.}, 
Phys. Rev. Lett. {\bf 62} (1989) 587; A. Schmid, Phys. Rev. Lett. 
{\bf 66} (1991) 80; F. von Oppen and E.K. Reidel, {\it ibid.} (1991) 84;
B.L. Altshuler {\it et al.}, {\it ibid.} (1991) 88

\bibitem{gefen}
        H.F. Cheung, E.K. Reidel, and Y.Gefen,
        Phys. Rev. Lett. {\bf 62} (1989) 587 
\bibitem{schwab}
        P. Schwab, Z. Phys. B {\bf 104} (1997) 97 

\bibitem{prl97} P. Mohanty, E.M.Q. Jariwala, R.A. Webb, Phys. Rev. Lett. {\bf 78}
(1997) 3366

\bibitem{prb97} P. Mohanty and R.A. Webb, Phys. Rev. B {\bf 55} (1997) R13452

\bibitem{levy2}
        L.P. Levy, Physica B {\bf 169} (1991) 245  

\bibitem{aak} B.L. Altshuler {\it et al.}, J. Phys. C{\bf 15} (1982) 7367

\bibitem{landau}
        E.M. Lifshitz and L.P. Pitaevskii, {\it Statistical Physics, Part 2},
        3rd ed., Pergamon Press, Oxford 1980

\bibitem{kravtsov} V.E. Kravtsov and V.I. Yudson, Phys. Rev. Lett. {\bf 70} (1993) 210;
		   A.G. Aronov and V.E. Kravtsov, Phys. Rev. B 47 (1993) 13409

\bibitem{imry} Y. Imry, H. Fukuyama and P. Schwab, cond-mat/9903017

\bibitem{ralph} A. Zawadowski, J. von Delft and D.C. Ralph, cond-mat/9902176

\bibitem{chakravarty}
        S. Chakravarty and A. Schmid, Phys. Rep. {\bf 140} (1986) 193


\bibitem{falko} V.I. Fal'ko and D.E. Khmel'nitskii, Sov. Phys. JETP {\bf 68} (1989) 186;
V. Fal'ko, Europhys. Lett. {\bf 8} (1989) 785; A.A. Bykov {\it et al.}, JETP Lett. {\bf 58} (1993) 543

\end{references}
\end{document}